\documentclass[prl,twocolumn,preprintnumbers,amsmath,amssymb]{revtex4}
\usepackage{graphicx}

\usepackage{dcolumn}% Align table columns on decimal point
\usepackage{bm}% bold math

\begin{document}

\title{From Inflation to Dark Energy}

\author{Robert Brout}
\affiliation{Facult{\'e} des Sciences, Universit{\'e} Libre de
Bruxelles, Belgium\\
and Perimeter Institute for Theoretical Physics\\
Waterloo, Ontario, Canada}

%\date{\today}

\begin{abstract}
It is proposed that after the macroscopic fluctuation of energy
density that is responsible for inflation dies away, a class of
microscopic fluctuations, always present, survives to give the
present day dark energy. This latter is simply a reinterpretation
of the causet mechanism of Ahmed, Dodelson, Green and Sorkin,
wherein the emergence of space is dropped but only energy
considerations are maintained. At postinflation times, energy is
exchanged between the "cisplanckian" cosmos and an unknown
foam-like transplanckian reservoir. Whereas during inflation, the
energy flows only from the latter to the former after inflation it
fluctuates in sign thereby accounting for the tiny effective
cosmological constant that seems to account for dark energy.

\end{abstract}

%\pacs{}

\maketitle

It has become apparent in recent years that the most important
component of energy that drives the Hubble expansion is the
so-called dark energy,    $(70\%)$.   It is homogeneously
distributed as opposed to cold dark matter, which is associated
with the galaxies.  Moreover its density remains relatively
constant in a range of $z$ values of $O(1)$. For these reasons
most astrophysicists and cosmologists consider dark energy to have
the characteristics of a cosmological constant, probably
associated with some form of vacuum energy.

We are then presented with the seemingly enigmatic situation that,
whereas one appeals to a phenomenological  $\Lambda$   to explain
cosmogenesis and inflation, of planckian origin and magnitude,
\cite{broutstaro,kolbturner},  at present we are confronted with
$\Lambda$ of a vastly different order of magnitude $(<10^{-100})$.
Do these phenomenologies have anything to do with each other?  In
this paper we argue in favor of the hypothesis that they are in
fact strongly related.  The present value of $\Lambda$  could well
be a remnant of inflation that is induced by fluctuations of
vacuum energy. After the macroscopic fluctuation, that is
responsible for inflation,  decays into quanta, \cite{kolbturner},
there remain fluctuations of vacuum energy in the presence of
these quanta (present day particles).  We reinterpret the causet
scenario of \cite{ahmed} in this light.

Our paper is laid out as follows: We first summarize the causet
mechanism \cite{ahmed} in preparation for its reinterpretation and
then make contact with inflation and our previous concept of the
inflaton.  We will also make contact with the creation of
cisplanckian modes, which one supposes must occur to keep
constant,  in the mean, the cisplanckian mode density ( equivalent
to a constant planckian cut-off), see \cite{kempfjacobson}. The
analogous scenario of mode creation in the black hole evaporation
process is discussed as well.

When the inflationary epoch is terminated, our interpretation of
the causet mechanism does not modify the quantitative realization
of \cite{ahmed} at least at this stage of development.  This
latter is given in terms of the parameter, $\alpha$  (see Eq.1
below), the value of which is rather tightly constrained on
phenomenological grounds \cite{ahmed}.  This parameter plays a
role in the estimate of the inflaton mass, using notions taken
from \cite{brout}.

It is evident, that like all phenomenologies which involve quantum
gravity and planckian physics, this paper is conjectural in
character. Its interest is to show that it is possible to
formulate hypotheses that unify cosmogenesis, inflation, dark
energy and mode creation.

A brief resume of \cite{ahmed} is now sketched.  Random
fluctuations of planckian units of energy (better,  of action) are
sequentially fed into the expanding cosmos.  In ref. 3 these lumps
are taken to be planckian elements of space -time as well. They
are distributed at random with a mean planckian density.  A
stochastic process is set up in time in which the events in the
i'th slice of time generates those in the $( i + 1)$th  [Eq.1
below].  This set of events is called a causet. Its observable
effects are limited to the past light cone of a comoving observer
hence proportional to the space time volume $(V)$ of the light
cone bounded sufficiently far back in the distant past so as to
validate the estimate $V \propto H^{-4}$ where  $H( t)$  is the
Hubble constant at the proper time of the observer. Two essential
assumptions are:

1. Designating  $\rho_M$  as the energy density  of everything
else but the vacuum energy density $(\rho_\Lambda)$, it is assumed
that as $\rho_M\rightarrow 0$ then $\rho_\Lambda \rightarrow 0$.
This is tantamount to the assumption that there is a stable
quiescent vacuum towards which the universe evolves at late times;
complete rest is the ``natural" condition of the cosmos. This
assumption is natural in that, were the vacuum state $\rho_M=0$
characterized by nonvanishing $\rho_\Lambda$, the universe in
which this vacuum exists would expand,  thereby creating energy,
$\rho_M$, which is nonvanishing due to the temporal dependence of
the metric.

For this very reason it is necessary to modify the first
assumption somewhat.  The final target state is metastable.  We
assume, as in \cite{kolbturner}, that large scale fluctuations,
those necessary to be seized upon by the cosmological component of
gravity to inflate and seed a new universe, are very rare.

2. Since $\Lambda$  varies in time (and in a more detailed
formulation it would vary in both space and time), there would be
sources of energy-momentum in general relativity which compensate
for this variation.  These would appear in cosmology in the
equation for the acceleration of $a$ ($a =$ scale factor). This
equation is dropped in \cite{ahmed}  and will be dropped in the
foregoing as well.

On the contrary the energy constraint (Friedmann equation) is
maintained, an equation of energy balance  between energy carried
by gravity due to the expansion and energy sources due to other
fields both in vacuum and in quanta.  This equation is a result of
invariance under time reparametrization, a physical principle one
would loathe to give up.

We remark here on the general point of view given in \cite{ahmed}
and which we shall adopt in the present work.  The system treated
is open. In \cite{ahmed} the causet elements arise from
 ``nowhere" but are created causally (sequentially) from
 those already present.
 ``Nowhere" can also be taken to mean from an unknown
 somewhere and this is the substance of our reinterpretation.
 The parameter $\alpha$
  is then interpretable as the squared matrix element
  of the transition from this
  ``somewhere" to the cosmos accompanied by energy exchange
  between the two.  This is what allows us to express the
   inflaton mass in terms of $\alpha$.

Our interpretation of \cite{ahmed} is perhaps more modest,
seemingly less audacious than the creation of space time lumps in
its quest for quantum gravity.  We suppose that space-time is
preexistent. The planckian lumps are the localized scenes where
energy exchange occurs between the transplanckian world (i.e.
degrees of freedom whose length scale is less than planckian,
often called space time foam) and the cisplanckian world, our
cosmos, the quanta therein and the vacuum in which these quanta
are situated.  This point of view allows one to conceive of
inflation, mode creation and the present day  $\Lambda$ on a
common basis.

It is commonly believed that the modes used to quantize fields in
usual (cisplanckian) quantum field theory supply a valid
description on scales of momenta  less than planckian.  $ (p < 1)$
where we use planckian units.  Herein there is an inherent frame
dependence taken to be the cosmological rest frame, a weakness one
hopes to eliminate in a future covariant formulation if such is
possible. For $p>1$, gravitational interactions among the modes
become the dominant component of energy.  One expects these modes
to become overdamped and field theory on this scale must be
considered a strong coupling theory.  A possibility that is often
envisaged is that fields on these  short length scales fold up
into localized structures of planckian and transplanckian
dimensions  and densities, Wheeler's so-called  foam.

The foam, in this image, serves as a reservoir of cisplanckian
modes which we now explain.  As the universe expands, wavelengths
expand with it. Therefore a cut off of modes at $p=O (1)$  at
earlier times, say $t = t_0$, becomes reduced by a factor $(a
(t_0) / a (t))$.

From inflationary times to the present  this can result in a
reduction in mode density $O(10^{-150})$.  But it is natural to
suppose that cisplanckian physics does not vary sensibly in time
in which case modes must be produced constantly.  One possibility
is that space is produced from ``nowhere" as in causet theory (see
\cite{ahmed})  and, in that space, modes of density $O (1)$ exist
because of the discretization. Another possibility is that space
is preexistent.  This is the hypothesis that has been explored in
\cite{brout} to explain the phenomenological inflaton and its mass
at the same time as mode creation.  In this scenario the degrees
of freedom that are locked into the localized transplanckian
structures are steadily solicited so as to maintain the
cisplanckian mode density constant as the universe expands. Vacuum
is conceived as a sort of state of phase equilibrium between two
fluids, the transplanckian reservoir and the cisplanckian world.
There must always be as many degrees of freedom in the reservoir
as is necessary to maintain this situation.  An example would be
supplied by the massy and/or short wave length modes of string
theory if ever that hypothesis could be made to work. In fact,
\cite{damour} supplies an image of high energy strings connected
to black hole structures which could serve as a nice model of foam
along these lines.

This cis-trans equilibrium thus is implemented by the exchange of
degrees of freedom from planckian lumps of energy to and from
modes. Localized structures of foam give and take energy from the
mean situation wherein the cisplanckian vacuum energy density is
maintained constant in the mean.   As in any equilibrium, there
will be fluctuations about this mean.  These will carry both signs
of energy and a sensible stochastic hypothesis for these
fluctuations could be Eq.1 wherein the fluctuation of action at
time $t = t_{i+ 1}$ is preconditioned by its value at time $t=
t_i$. Independently of any detailed mechanisms, as pointed out in
ref 3, the order of magnitude of vacuum energy density at time
$t$, in a universe where matter is present, is $\pm [H(t)]^2$
since the volume of the backward light cone is  $O[(H(t)^{ -4}]$
and the net energy transfer in this light cone is $\pm N
(t)^{1/2}$ where $N (t)$ is the number of cis-trans exchanges
which are supposed to be planckian in energy as well as density.
Thus, $N(t)$ is $O [(H(t)^{ -4}]$  as well, whence
$\rho_\Lambda\propto O ( N^{1/2} / H^{-4}) = H^2 (t)$. As stated,
our reinterpretation maintains this remarkably simple and
successful explanation of $\rho_\Lambda$.

There is one very important point that must be emphasized, to wit:
it is supposed that only the cisplanckian energy fluctuations
result in the metric back reaction of general relativity here
given by the variation for the Hubble constant appearing in the
energy constraint.  Degrees of freedom locked into the
transplanckian structures do not have the long wavelength
components available to affect a macroscopic back reaction. Indeed
it is not even clear that Einsteinian general relativity as we
usually practice it applies at all to the transplanckian world.
One guesses it would not.

We may understand the above statement in a transparent way by
considering theories of black hole evaporation that tame the
transplanckian problem through mode production or less
dramatically through conversion of over damped to under damped
degrees of freedom (\cite{broutspindel,parentani}).

When one tracks an evaporated Hawking photon backwards in time one
finds that its trajectory hugs the horizon at exponentially short
lengths hence describing a vacuum fluctuation in the Schwarzschild
metric which has exponentially large proper energy (eigen value of
$\delta/\delta r $, $r$ being the proper radial coordinate). Thus
the description in terms of modes fails in this region. Rather one
must suppose that there is a planckian skin containing
transplanckian structures about the horizon which feeds into the
mode description once the fluctuation of the outgoing
configuration ( in a wave packet description) is outside the skin.
Then by subsequent Bogoljubov transformation, this fluctuation in
part becomes an outgoing quantum and the energy constraint results
in a back reaction of the metric described by a loss of black hole
mass given by the frequency of that quantum in the asymptotic
region.   The transplanckian reservoir serves only to feed the
cisplanckian modes and the metric back reaction is the back
reaction to this latter part only.   We suppose that something of
this sort also happens in the cosmological expansion wherein the
cisplanckian field configuration arises out of the transplanckian
structure due to expansion.

On the basis of this example, one may develop a predilection in
favour of our interpretation that causets do not concern the
creation of space in lumps, but only the exchange of energy
associated with these lumps.   Indeed, in the black hole example,
space is always present or presumed as such in all existing
models.  The transplanckian skin about the horizon in the type of
model we are discussing is the locus of sites of cis trans
exchange.  Note that if it turns out that the correct black hole
transplanckian taming is in-out scattering, as advocated for
example by G. 't Hooft, then this example will be irrelevant.

Let us  now turn to inflation.  The parameter $\alpha$   in
\cite{ahmed} is defined through the recursion relation Eq.5 in
\cite{ahmed}.
\begin{equation}
\rho_{\Lambda,i+1}=\frac{S_{i+1}}{N_{i+1}}=\frac{S_{i}+\alpha
\xi_{i+1}(\delta N_i)^{1/2}}{N_i+\delta N_i}
\end{equation}

    The index $i$ denotes the i'th
    slice through the backward light cone at time
    $ti$, $N_i$  is the total number
    of lumps accumulated at time $t_i$
    in that cone.  $\xi_i$   is a  random variable of mean
    zero and standard deviation $1$.  $\delta N_i$
    is the number of lumps in the ith  slice.
    $\alpha$  is a numerical parameter.  Good agreement with
    phenomenology was found in \cite{ahmed}  with   $\alpha=O(10^{-2})$.
     See \cite{ahmed} for further detail and the understanding why  $\alpha$
      must be constrained to $O(10^{-2})$.
      It is no small achievement that there exist runs which seem
      to be accurate replicas of phenomenology back to $z$
       values of $O(1)$.

We  have presented  these details to motivate our interpretation
of  $\alpha$  as the squared  matrix element  for cis trans
exchange of a planckian unit of energy between the transplanckian
lump in the reservoir and the cisplanckian world.

The relevance for inflation and the mass of the inflaton is now
presented. As in Refs.\cite{broutstaro, kolbturner}, we start at
some time when space is flat hence $\rho_M=0$, and $\rho_\Lambda$,
in the mean, is $0$ as well. Consistent with the assumption of our
target values as discussed above.  this vacuum state is taken to
be an equilibrium between the cis and trans fluids.

Though small scale length fluctuations regress, there exist large
scale fluctuations which will inflate. (See \cite{kolbturner}). It
was argued in \cite{kolbturner} that this large fluctuation is the
phenomenological inflaton often represented as a scalar field but
which is commonly believed to be a manifestation of quantum
gravity.
  In the usual inflationary scenarios this macroscopic
  fluctuation will slowly regress after the number of e-folds is
  sufficient to account for our observed universe.
  The  initial fluctuation is chosen sufficiently large to do the
  job \cite{kolbturner}.
Of course, small scale fluctuations accompany this regression.  In
current phenomenology one accounts for the spectrum of cosmic
background radiation in terms of one class of these fluctuations.
But we now are appealing to another class, those that arise
explicitly from the fluctuations of the cis trans exchange
equilibrium and which is our interpretation of \cite{ahmed}.  The
big difference between the physics of the macroscopic fluctuation
responsible for inflation and the microscopic fluctuations which
have been discussed until now is that these latter occur in a
universe already filled with energy.   As a result, the
cisplanckian system can both take on and give up energy in these
exchanges.  This is the natural and beautiful explanation of
\cite{ahmed} for why $\rho_\Lambda = O(H^2)$  at present.  On the
contrary, the macroscopic fluctuation called the inflaton is a
phenomenon which takes place when  $\rho_M=0$.  The cisplanckian
system apart from microscopic fluctuations then has but negligible
energy to give up during the evolution of the large macroscopic
fluctuation. Almost all of the transplanckian lumps within this
large fluctuation only give energy to the cisplanckian degrees of
freedom i.e. the modes, thereby creating our universe after
reheating.  As we have emphasized the macroscopic metric back
reaction encoded in the expanding scale factor is due only to
changes in energy in the cisplanckian world.

We thus understand why the inflationary Hubble constant,  or
equivalently the value of  $\rho_\Lambda$     during inflation  is
of planckian scale.  All the transplanckian lumps that give up
energy act in consort  to increase the total energy of the
inflating universe until towards the end when the inflatonic motor
runs out; preheating takes place and the isentropic expansion
ensues. Thus the energy density induced in the cisplanckian world
at inflationary times is no longer $O(N^{1/2}/ V)$  [where we
recall that $N$ is the number of lumps that affect the trans cis
exchange and V is the space time volume of the backward light
cone]. Rather in the inflationary epoch the energy density passed
to the cisplanckian world is $O (N/V) ~[= (H^{-4}/H^{-4}) =O(1)]$.
In short, inflation is a process of coherent passage to the
cisplanckian world of energy from transplanckian lumps whereas the
dark energy is a random incoherent process with either sign of
energy exchange.  At the same time, and presumably by the same
mechanism cisplanckian modes are created from the transplanckian
degrees of freedom.

One may refine somewhat the estimate of the order of magnitude of
$\rho_\Lambda$ during inflation or at least the effective inflaton
mass by appeal to $\alpha$ and its interpretation as the squared
matrix element for the exchange.

It was argued in \cite{brout} that the inflaton is a wave of
varying density induced by fluctuations from the equilibrium
resulting from cis trans exchanges.  On this basis the inflaton
mass was estimated from the equation
\begin{equation}
\mu^2=pM^2
\end{equation}
where $p =$  probability of cistrans exchange and M equal mass
scale of the matrix element for the exchange.  We here identify $p
=\alpha ~~(= O (10^{-2}))$.   Phenomenologically, $\mu=
      O (10^{ -6} - 10^{-5} )$ whence $M = O (10^{ -4})$,
      a not unreasonable number but nevertheless rather lower
      than what one might have expected.  Only when a fundamental
      theory of planckian and transplanckian physics begins to
      take on a credible form will we be in a position to
      confirm or infirm these considerations.

After the first draft of this article I had occasion to hear
Renaud Parentani present his recent work on the production of cis
planckian modes from a transplanckian reservoir field.  The
essential input of his theory is a set of stochastic assumptions
governing the reservoir field, designed to give the desired result
of a mean planckian density of cis planckian modes during the
Hubble expansion. The output is the elegant passage from over to
under damping of degrees of freedom at the planckian level.  It is
clear that this point of view has much in common with ideas
expressed in the above paragraphs and it will be most interesting
to see if this recent work can furnish more quantitative detail on
our point of view of dark energy and inflation.

\bf Acknowledgement: \rm It is a pleasure for me to acknowledge
with gratitude the friendly patient explanations of Raphael
Sorkin, of the beautiful ideas expressed in \cite{ahmed}.  During
this period I was hosted by the Perimeter Institute, Waterloo,
Ontario, Canada.  I wish to express my gratitude to the many
people who helped me at Perimeter and my appreciation of the
productive and friendly atmosphere that prevails there; in
particular, Achim Kempf has been most helpful and kind.

\vfill

\begin{thebibliography}{99}
% superconductivity
\bibitem{broutstaro} R. Brout, F. Englert , E. Gunzig, Gen. Rel.
Grav. {\bf 10} 1 (1979) (1st Prize Gravity Award Essay), A.A.
Starobinski, Phys. Lett. {\b 91B} 99 (1980)

\bibitem{kolbturner} E.W. Kolb, M.S. Turner, \it The Early Universe, \rm
Perseus (1993)

\bibitem{ahmed} M. Ahmed, S. Dodelson, P.B. Greene, R. Sorkin,
Phys. Rev. {\bf D69} 103523 (2004)

\bibitem{kempfjacobson} A. Kempf, Phys. Rev. {\bf D63} 083514
(2001), B. Z. Foster, T. Jacobson, JHEP 0408:024 (2004)

\bibitem{brout} R. Brout, gr-qc/0201060

\bibitem{damour} T. Damour, G. Veneziano, Nucl. Phys. {\bf B568}
93 (2000)


\bibitem{broutspindel} R. Brout, C. Gabriel, M. Lubo, P. Spindel,
Phys. Rev. {\bf D59} 044005 (1999)

\bibitem{parentani} C. Barrabes, V. P. Frolov, R. Parentani, Phys. Rev.
{\bf D59} 124010 (1999), Phys. Rev. {\bf D62} 044020 (2000)


\end{thebibliography}
\end{document}